# Observation of non-Hermitian degeneracies in a chaotic exciton-polariton billiard

T. Gao[1], E. Estrecho[1], K.Y. Bliokh[1,2], T.C.H. Liew[3], M.D. Fraser[2], S. Brodbeck[4], M. Kamp[4], C. Schneider[4], S. Höfling[4], Y. Yamamoto[5], F. Nori[2], Y.S. Kivshar[1], A. Truscott[1], R. Dall[1], and E.A. Ostrovskaya[1]

[1]Research School of Physics and Engineering, The Australian National University, Canberra, ACT 2601, Australia
[2]Center for Emergent Matter Science, RIKEN, Wako-shi, Saitama 351-0198, Japan
[3]School of Physical and Mathematical Sciences, Nanyang Technological University, 637371, Singapore
[4]Technische Physik and Wilhelm-Conrad-Röntgen Research Center for Complex Material Systems, Universität Würzburg, Am Hubland, D-97074 Würzburg, Germany
[5]ImPACT Project, Japan Science and Technology Agency, Chiyoda-ku, Tokyo 102-0076, Japan

**Exciton-polaritons are hybrid light-matter quasiparticles formed by strongly interacting photons and excitons (electron-hole pairs) in semiconductor microcavities [1–3]. They have emerged as a robust solid-state platform for next-generation optoelectronic applications as well as fundamental studies of quantum many-body physics. Importantly, exciton-polaritons are a profoundly *open* (i.e., non-Hermitian [4,5]) quantum system: it requires constant *pumping* of energy and continuously *decays* releasing coherent radiation. Thus, the exciton-polaritons always exist in a balanced potential landscape of *gain* and *loss*. However, the inherent *non-Hermitian nature* of this potential has so far been largely ignored in exciton-polariton physics. Here we demonstrate that non-Hermiticity dramatically modifies the structure of modes and spectral degeneracies in exciton-polariton systems, and, therefore, will affect their quantum transport, localisation, and dynamical properties [6–10]. Using a spatially-structured optical pump [11–13], we create a *chaotic exciton-polariton billiard*. Eigenmodes of this billiard exhibit multiple non-Hermitian spectral degeneracies – *exceptional points* [14,15]. These are known to cause remarkable wave phenomena, such as unidirectional transport [16,19], anomalous lasing/absorption [17,18], and chiral modes [20]. By varying parameters of the billiard, we observe *crossing* and *anti-crossing* of energy levels and reveal the nontrivial topological modal structure exclusive to non-Hermitian systems [9,14–24]. We also observe the mode switching and *topological Berry phase* for a parameter loop encircling the exceptional point [25,26]. Our findings pave the way for studies of non-Hermitian quantum dynamics of exciton-polaritons, which can lead to novel functionalities of polariton-based devices.**

Studies of open quantum systems go back to Gamow's theory of nuclear alpha-decay developed in the early days of quantum mechanics [4,27]. Indeed, metastable states of a single quantum particle in a spherically symmetric potential well with semi-transparent barriers decay in time, and therefore are characterized by *complex* energies. Furthermore, introducing a 2D potential well with nontrivial geometry, i.e., *quantum billiard*, results in strongly correlated energy levels and transition to *quantum chaos* [6,21,23,28–31]. *Spectral degeneracies* crucially determine transport and dynamical properties in both non-Hermitian and chaotic wave systems [6–10,16–19]. In chaotic and disordered wave systems, spectral degeneracies underpin statistical properties and quantum phase transitions from the localised to delocalised dynamics [7–10]. In non-Hermitian (including $\mathcal{PT}$-symmetric) systems, nontrivial topology of eigenmodes and unusual transport properties in the vicinity of *exceptional points* [16–19,21] are currently under intense investigation. To date, basic non-Hermitian or stochastic dynamics were studied in the context of microwave [9,20–22,26], optical [16–19,21,23], atomic [24,29,30,32], and electron [28,31,33] waves. However, the concepts of non-Hermiticity and quantum chaos remain largely separated from each other, due to the lack of a simple quantum system in which both features would be readily accessible. Moreover, it is challenging to produce *artificial complex potentials* with gain and loss for classical waves, as well as to observe *nanoscopic electron states* in solids.

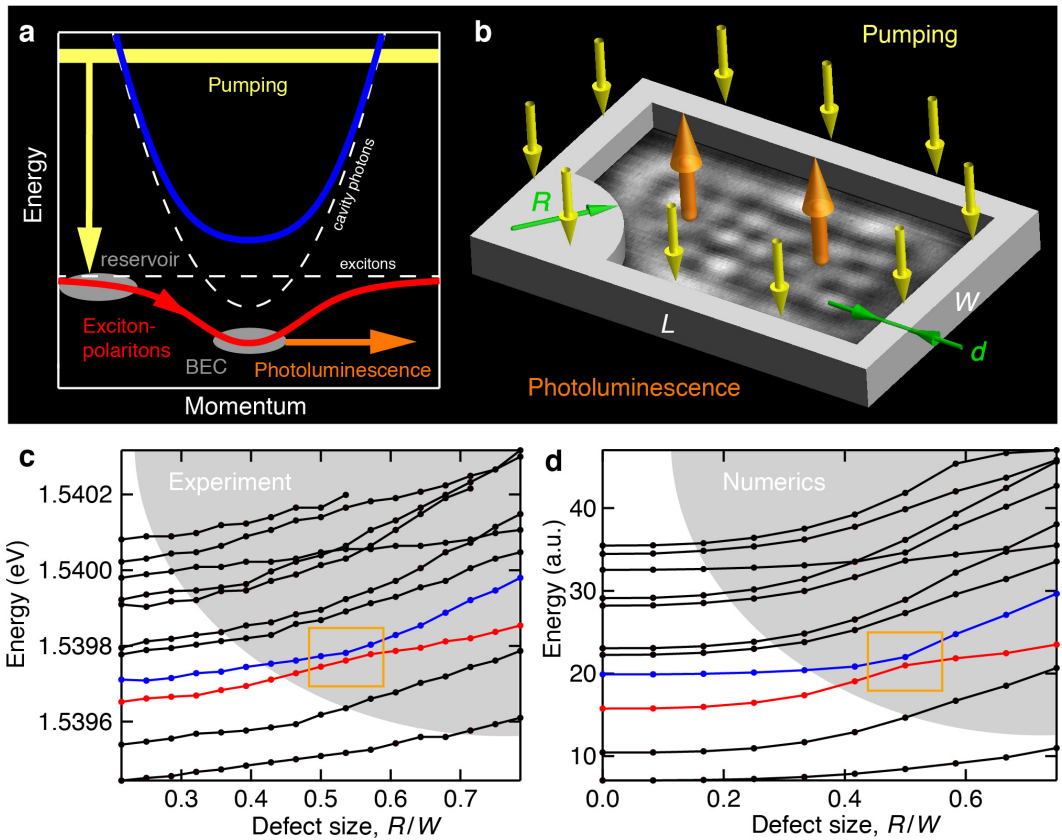

**Figure 1. Non-Hermitian exciton-polariton Sinai billiard and its spectrum. (a)** Exciton-polariton dispersion showing the incoherent excitonic reservoir, continuously replenished by the optical pump and polariton BEC decaying into cavity photoluminescence [1–3]. **(b)** Schematics of the exciton-polariton Sinai billiard formed in the plane of a quantum well embedded into the microcavity (see Supplementary Information). The barrier is induced by the optical pump via the exciton reservoir, and the wave function of the confined polariton BEC (shown in greyscale inside the billiard) is imaged via the photoluminescence. The billiard dimensions are $W = 14\,\mu m$, $L = 23\,\mu m$, the radius of the defect $R$ is varied from 0 to $W$, and the thickness of the walls $d$ is varied from $3\,\mu m$ to $7\,\mu m$ (see Supplementary Information). Experimentally measured **(c)** and numerically-simulated **(d)** spectra $E(R)$ for the first 11 modes of the billiard. With growing $R$, numerous degeneracies and quasi-degeneracies proliferate in the grey area, which signals the transition to quantum chaos [6]. Evolution and topological structure of two near-degenerate modes (red and blue in the yellow rectangle) are analysed in detail in Figs. 2–4.

Microcavity exciton-polaritons represent a unique quantum macroscopic system, which combines the main advantages of *light* and *matter* waves [1–3]. Being bosons, exciton-polaritons can display collective quantum behaviour, *Bose-Einstein condensation* (BEC), when they occupy a single-particle quantum state in massive numbers. Exciton-polaritons have provided a very accessible playground for studies of the collective quantum behaviour because they condense at the temperatures ranging from 10 K to *room temperature* (compared to nano-Kelvins for neutral atoms) and do not require painstaking isolation from the environment. The schematics of exciton-polariton condensation under continuous-wave incoherent optical excitation conditions [1] are shown in Fig. 1a. The optical pump, far detuned from the exciton resonance in the cavity, effectively creates an incoherent reservoir of 'hot', exciton-like polaritons. Above a threshold phase-space density of the reservoir, relaxation and stimulated scattering into the coherent BEC state of exciton-polaritons dominate the dynamics. The continuously pumped condensate decays releasing coherent photons, which escape the cavity carrying all information about the condensed state. The interactions between the reservoir and condensed exciton-polaritons are responsible for the formation of the *effective pump-induced potentials* [11–13]. Thus, the macroscopic *matter wave function* is shaped by an optical pump and spatially resolved via free-space *optical microscopy*. This enables us to clearly observe and control non-Hermitian and irregular quantum dynamics.

We use a structured optical pump [11–13] to create a non-Hermitian potential in the shape of a *chaotic Sinai billiard* [6] (see Fig. 1b) for condensed exciton-polaritons (see Supplementary Information for details). The radius of the circular defect in this billiard, $R$, controls the transition from the regular to chaotic dynamics [6]. In our experiment the billiard has 'soft' (inelastic) walls of a finite width and height. The main properties of eigenstates of the exciton-polariton condensate in the billiard can be described by a linear Schrodinger equation with a complex two-dimensional potential $V(\mathbf{r}) = V'(\mathbf{r}) + iV''(\mathbf{r})$. Here $V'(\mathbf{r}) \propto P(\mathbf{r})$ is the potential barrier shaped as a Sinai billiard boundary with a Gaussian envelope. This potential is proportional to the optical pump rate, $P(\mathbf{r})$, and is induced by the strong repulsive interaction between the excitonic reservoir populated by the pump and the polariton BEC [11–13]. The imaginary part of the potential, $V''(\mathbf{r}) \propto P(\mathbf{r}) - \gamma$, combines the gain profile produced by the same optical pump $P(\mathbf{r})$ with the spatially-uniform loss $\gamma$ due to the polariton decay (Fig. 1b). Despite the strong polariton-polariton interactions, the corresponding nonlinearity mostly affects the relative population of the energy eigenstates, as well as the overall blueshift (see Supplementary Information).

Figures 1c and 1d show the experimentally measured and numerically computed energy spectra $E(R)$ of the first 11 levels as a function of the defect radius, $R$. Changing the radius varies the geometry of the billiard and, hence, affects the energy levels. One can see that multiple degeneracies and near-degeneracies proliferate in the spectrum with growing *R*, signalling the transition from regular to chaotic dynamics [6]. In Hermitian billiards, the levels generically *avoid crossings (anti-cross)*, which correspond to the average level repulsion and Wigner distribution of the nearest-neighbor energy spacings [8]. In contrast, the non-Hermitian systems can exhibit both *crossings* and *anti-crossings* of levels [9,21–24]. This is because the energy eigenvalues in non-Hermitian systems are *complex*: the real part and imaginary parts correspond to the *real energies* and *linewidths* of the modes, respectively. A crossing of the energies is accompanied by an anti-crossing of the linewidths and vice versa. In our experiment, we measure the spectral profile of the cavity photoluminescence at a particular spatial position and extract both peak energies and widths of spectral resonances (see Supplementary Information). Crossings as well as anti-crossings of real energy levels are clearly seen both in experiments (Fig. 1b) and numerics (Fig. 1c).

To observe the transition between the crossing and anti-crossing for the *same* near-degenerate pair of eigenvalues, one needs to vary *a second* control parameter. In our exciton-polariton billiard this additional parameter is the thickness, $d$, of the billiard walls. Provided the internal area of the billiard remains intact, this parameter does not affect the geometry of the billiard and primarily controls the *imaginary part* $V''$ of the non-Hermitian potential barrier. Figure 2 shows one pair of billiard modes highlighted in Fig. 1b in the vicinity of a near-degeneracy for two values of the control parameter $d$. One can clearly see the anti-crossing (crossing) behaviour of the real (imaginary) parts of the complex eigenenergies in the billiard with thick walls (Figs. 2a,c) and the opposite behaviour for the thin-wall billiard (Figs. 2b,d).

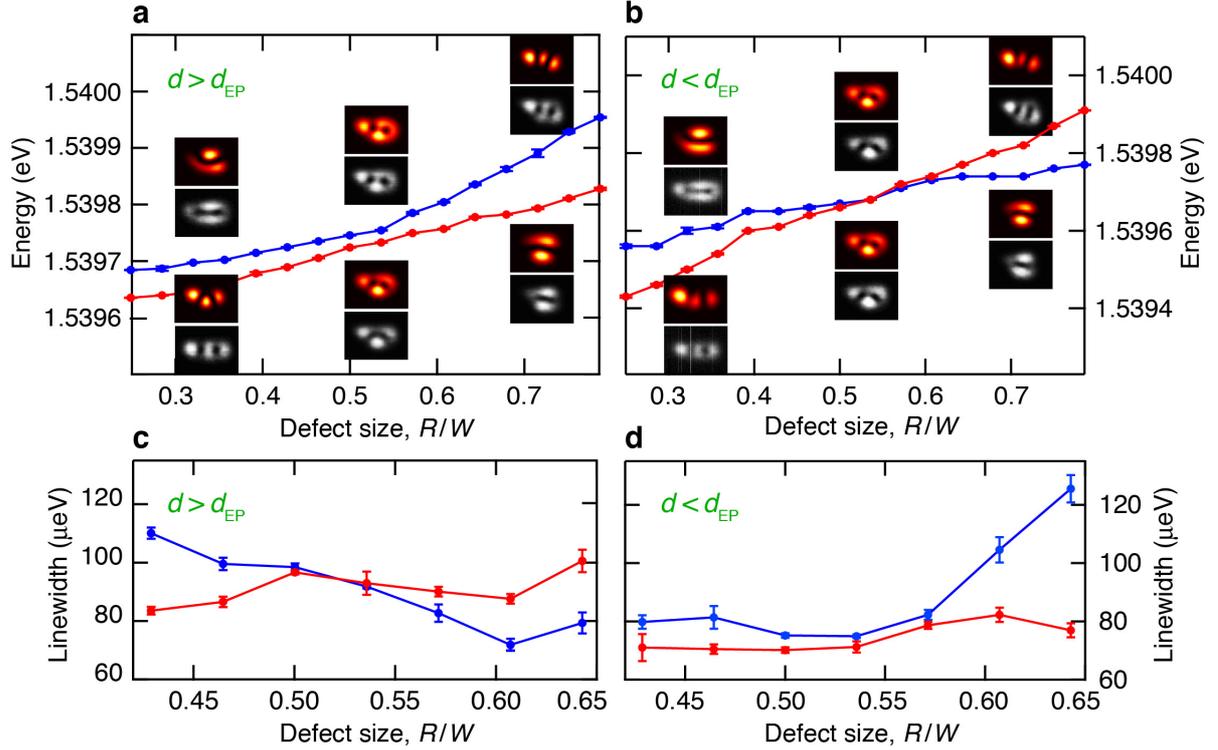

**Figure 2. Crossing and anti-crossing for the two near-degenerate modes highlighted in Figs. 1c,d.** Experimentally observed anti-crossing **(a)** and crossing **(b)** of eigenenergies of two modes in the spectrum of the exciton-polariton Sinai billiard with varying parameter $R$ (see Fig. 1) for thick, $d \simeq 6\,\mu m$, **(a,c)** and thin, $d \simeq 4\,\mu m$, **(b,d)** billiard walls. Panels **(c)** and **(d)** show the corresponding crossing and anti-crossing of the linewidths (i.e., imaginary parts of the complex eigenvalues). The upper (lower) inset panels in (a,b) illustrate the numerically calculated (experimentally imaged) spatial structure of the eigenmodes at different values of the parameter $R$. Details of the hybridisation region are found in the Supplementary Information.

Importantly, the energy-resolved real-space imaging of the photoluminescence provides all the information about complex eigenvalues as well as the *spatial structure of the eigenmodes* (wavefunctions). In particular, the levels shown in Figure 2 correspond (at $R=0$) to the third mode with three horizontal lobes and the forth mode with two vertical lobes. The experimentally imaged and calculated spatial profiles of these eigenmodes are shown as insets in Figures 2a,b along the eigenenergy curves. We observe that the two modes are *hybridised* and therefore change their spatial profiles in the near-degeneracy region, and 'exchange' their spatial profiles after passing it.

The behaviour of two billiard modes in the vicinity of a degeneracy can be described by a simple model of a two-level system with an effective coupling (see the Supplementary Information). The corresponding non-Hermitian Hamiltonian reads [9,20–24]:

$$\hat{H} = \begin{pmatrix} \tilde{E}_1 & q \\ q^* & \tilde{E}_2 \end{pmatrix}, \quad \tilde{E}_{1,2} = E_{1,2} - i\Gamma_{1,2}. \tag{1}$$

Here $\tilde{E}_{1,2}$ are the complex eigenvalues of two uncoupled modes (with $E_{1,2}$ being the real energies and $\Gamma_{1,2}$ being the decay/gain rates), whereas $q$ characterizes the coupling between these two modes. We will also use the mean complex energy $\tilde{E}=(\tilde{E}_1+\tilde{E}_2)/2\equiv E-i\Gamma$, and the complex energy difference $\delta\tilde{E}=(\tilde{E}_2-\tilde{E}_1)/2\equiv \delta E-i\delta\Gamma$. The eigenvalues of the Hamiltonian (1) are $\lambda_{1,2}=\tilde{E}\pm\sqrt{\delta\tilde{E}^2+|q|^2}$; their real and imaginary parts, which depend on the parameters $\delta\tilde{E}=(\delta E,\delta\Gamma)$, are shown in Figure 3. These complex eigenvalues coalesce, $\lambda_1=\lambda_2$, in the *exceptional points* (EPs) [14–24], where $i\delta\tilde{E}_{EP}=\pm|q|$. At this points, the eigenstates also coalesce and form a single chiral mode [14,15,20]. Assuming that the coupling constant $q$ is fixed, the EPs appear in the parameter plane as $(\delta E_{EP},\delta\Gamma_{EP})=(0,\pm|q|)$. We assume $\delta\Gamma>0$ in our range of parameters, so that there is only one exceptional point in the domain of interest. The EP can be encircled in the $(\delta E,\delta\Gamma)$ plane by varying these two parameters, as seen in Fig. 3.

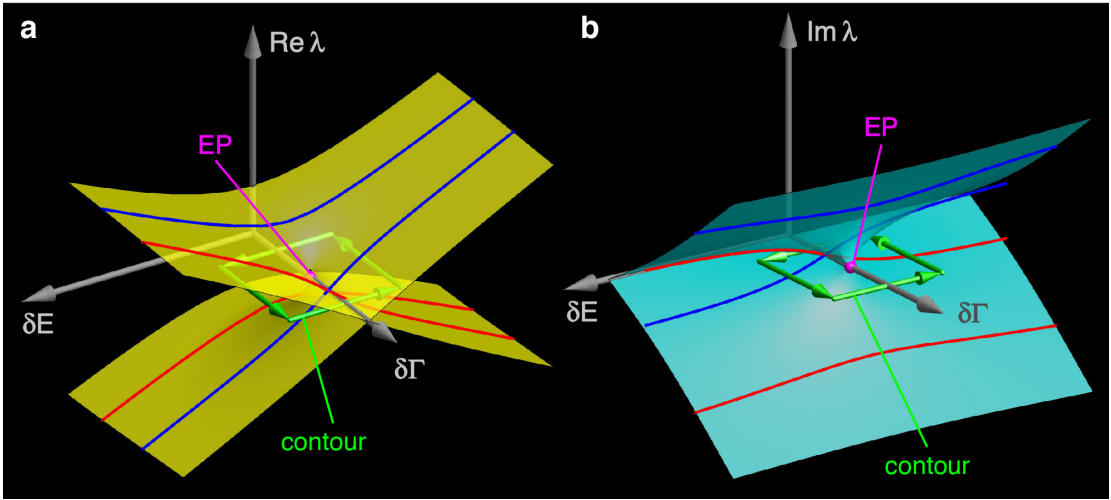

**Figure 3. Eigenvalues of a two-level non-Hermitian model in the vicinity of the exceptional point.** Real **(a)** and imaginary **(b)** parts of the eigenvalues $\lambda_{1,2}$ of the model (1) as functions of two parameters $\delta E$ and $\delta\Gamma$. The exceptional point (EP) is shown in magenta. The crossing and anti-crossing of the real and imaginary parts of the eigenvalues as functions of $\delta E$, for $\delta\Gamma<\delta\Gamma_{EP}$ and $\delta\Gamma>\delta\Gamma_{EP}$, are shown in red and blue. This is in correspondence with the experimentally observed behaviour in Fig. 2. The traverse along the green contour encircling the exceptional point in the $(\delta E,\delta\Gamma)$ plane reveals the nontrivial topology of eigenmodes, as shown in Fig. 4.

Two parameters of the model, $(\delta E,\delta\Gamma)$, approximately correspond to the varying parameters $(R,d)$ of our exciton-polariton billiard. The radius $R$ mostly affects the energy difference between the modes, while the thickness $d$ of the billiard walls controls the gain/loss balance between the modes. Note that different modes have different spatial overlaps with the imaginary potential $V''(\mathbf{r})$, and, therefore, are characterised by different integral (spatially-averaged) dissipation parameters $\Gamma_{1,2}$ (see the Supplementary Information). In our case, increasing $R$ corresponds to increasing $\delta E$, while increasing $d$ corresponds to decreasing $\delta\Gamma$. The effective coupling $q$ is determined by the spatial overlap between the two modes away from the hybridisation region [9]. The red and blue curves in Figure 3 show the crossing/anti-crossing behaviour of the real and imaginary parts of the eigenvalues versus the energy difference $\delta E$ for two values of the dissipation parameter: $\delta\Gamma<\delta\Gamma_{EP}$ and $\delta\Gamma>\delta\Gamma_{EP}$. This behaviour is perfectly consistent with that in the experimental Figure 2, which means that our range of varying parameters includes the exceptional point.

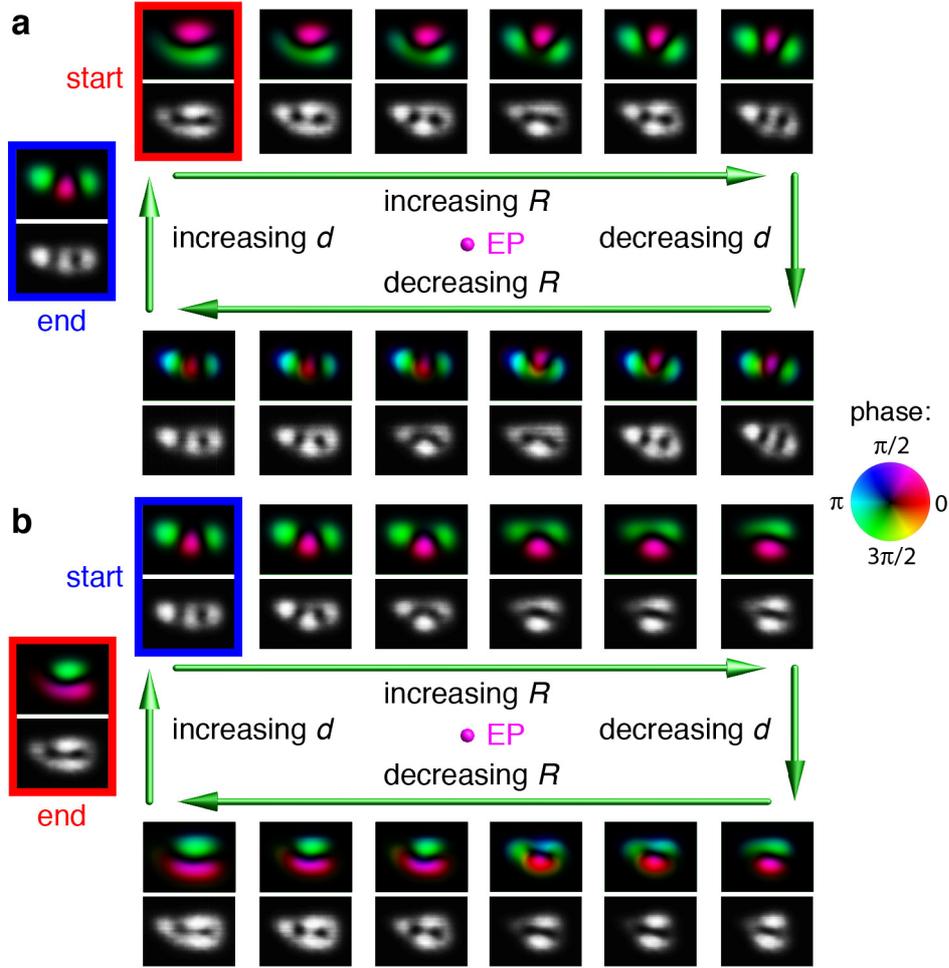

**Figure 4. Observation of the topological Berry phase for circling around the exceptional point.** Evolution of spatial distributions of the selected mode (from the pair shown in Fig. 2) along the closed contour in the parameter space $(R,d) \sim (\delta E, \delta \Gamma)$ encircling the exceptional point (see Fig. 3). The first loop **(a)** shows the transition to a different branch (mode) through the hybridization region (see explanations in text). The second loop **(b)** returns the mode to the original one with a $\pi$ *topological phase* shift [25,26]. The phases are inferred from comparison with the numerically calculated modes.

The complex-eigenvalues structure in the vicinity of the exceptional point reveals nontrivial topology of a branch-point type [14–24] shown in Figure 3. Therefore, continuous encircling of the non-Hermitian degeneracy in the two-parameter plane (e.g., along the green contour in Fig. 3) results in the transition to the other branch. When the contour is traversed *twice*, we return to the original mode, most significantly with a $\pi$ *topological phase* shift [25,26]. We used the method suggested in the microwave experiment [26] to trace the topological evolution of two modes around the exceptional point. Figure 4 depicts the experimentally measured intensities and the corresponding numerically simulated phase profiles of the two modes (from Fig. 2) evolving along the contour in the $(\delta E, \delta \Gamma) \sim (R,d)$ plane (see Fig. 3) and encircling the exceptional point. Namely, in Figure 4a, we start on the upper branch (blue in Figs. 2a and 3a) at $R < R_{EP}$, $d > d_{EP}$ and increase the radius to $R > R_{EP}$. This takes us from the vertical two-lobe mode, through the anti-crossing, to the horizontal three-lobe mode (still on the blue level). Then, we decrease the thickness to $d < d_{EP}$ and stay on the same horizontal three-lobe mode, which now corresponds to the *red* branch in Figs. 2b and 3a. Next, reducing the radius *R* takes this mode through the crossing and recovers its three-lobe structure. Increasing *d* closes the loop. Thus, the continuous evolution brought us from the vertical two-lobe mode ('start' in Figure 4a) to the horizontal three-lobe mode ('end' in Figure 4a) at the same values of parameters. Repeating this traverse one more time (Figure 4b) returns us to the original vertical two-lobe mode, but now with the $\pi$ *topological phase* shift (clearly seen in the simulated phase profiles). The experimental density distribution of the modes is in very good agreement with that calculated numerically. Therefore we can associate *the phase structure* of the simulated spatial modes with the experimental mode profiles (cf. [25]).

Thus, we have demonstrated a chaotic non-Hermitian exciton-polariton billiard with multiple spectral degeneracies. We have provided detailed experimental observations of the non-trivial behaviour of complex eigenvalues and eigenmodes in the vicinity of an exceptional point. These include crossing/anti-crossing transitions as well as mode switching and topological Berry phase for the cyclic evolution around the exceptional point in the two-parameter plane. Our results show that the inherent non-Hermitian nature of exciton-polaritons determines their basic properties, which are crucial for transport and quantum information processing. Therefore, these features should be taken into account in future studies and applications involving confinement and manipulation of exciton-polaritons. Most importantly, this complex quantum dynamics can bring novel functionality to polariton-based devices operating at the interface between photonics and electronics. Generally, exciton-polaritons offer a novel macroscopic quantum platform for studies of non-Hermitian physics and quantum chaos at the interface between light and matter.

**Acknowledgements.** We are grateful to Prof. Michael Berry for insightful comments. This research was supported by the Australian Research Council, the ImPACT Program of Council for Science, Technology and Innovation (Cabinet Office, Government of Japan), the State of Bavaria, RIKEN iTHES Project, MURI Center for Dynamic Magneto-Optics, and Grant-in-Aid for Scientific Research (S).